\documentstyle[epsfig]{jaa}
\input{def.lst}
\volume{20}
\pubyear{1999}
\pagerange{\pageref{firstpage}---\pageref{lastpage}}
\begin{document}
\setcounter{page}{0}
\label{firstpage}
\title[The interstellar clouds of Adams and Blaauw - II]{The interstellar clouds of Adams and Blaauw revisited: an HI absorption study - II}

\author[Rajagopal, Srinivasan \& Dwarakanath]{Jayadev Rajagopal, G.Srinivasan and K.S.Dwarakanath\\Raman Research Institute,\\ Bangalore 560 080, India.\\email: jayadev@stsci.edu, srini@rri.ernet.in, dwaraka@rri.ernet.in}

\date{}
  \maketitle

\begin{abstract}
In the preceding paper (paper I), we presented HI absorption
spectra towards radio sources very close to the lines of sight
towards 25 bright stars against which optical absorption spectra
had been obtained earlier. In this paper we analyse the results
and draw some conclusions.

To summarize briefly, in most cases we found HI absorption
at velocities corresponding to the optical absorption features
provided one restricted oneself to velocities $\lsim$ 10 km s$^{-1}$.
At higher velocities we did not detect any HI absorption down to an
optical depth limit of 0.1 (except in four cases which we attribute
to gas in systematic motion rather than clouds in random motion).
After discussing various scenarios, we suggest that this trend should
perhaps be understood in terms of the high velocity interstellar clouds
being accelerated, heated and ablated by expanding supernova remnants.
\end{abstract}

\section{Introduction}
In the preceding paper (Rajagopal, Srinivasan and Dwarakanath, 1998; paper I)
we presented the results of a program to obtain the HI absorption profiles
towards a selected sample of bright stars. As mentioned there, detailed 
optical absorption studies exist in the direction of these stars in the
lines of NaI and CaII.
The motivation for such an observational program was also described in 
the previous paper. In the present paper, we wish to discuss the results obtained
by us. There are two major issues that we wish to address and discuss later:
\begin{itemize}
\item[1] Are the properties of interstellar clouds seen in optical absorption
the same as those seen in HI emission and absorption in general?
\item[2] What is the origin and nature of the faster clouds seen in optical
and UV absorption ?
\end{itemize}
Let us elaborate a bit on the first issue. Although the general picture of
the ISM that emerged from optical and radio observations, respectively, is
the same, viz., clouds in pressure equilibrium with an intercloud medium, it
has not been possible to directly compare the inferred properties. Whereas
both the column densities and the spin temperatures of the clouds have
been estimated from HI observations, there have only been indirect and
often unreliable estimates for the clouds seen in optical absorption.
Our HI absorption measurements, combined with HI emission measurements
in the same directions will enable us to directly estimate for the first
time the column densities and spin temperatures of the clouds seen in
optical absorption.

The second question mentioned above arises as follows. The existence
of a {\it high velocity tail} in the distribution of random velocities 
of clouds was firmly established by Blaauw (1952) from the data obtained
by Adams (1949). As mentioned in paper I, it was noticed quite early on
by Routly and Spitzer (1952) that the faster clouds have a smaller NaI to
CaII ratio than the lower velocity clouds. Early HI emission measurements
(see paper I for references) in the direction of the bright O and B stars 
provided an added twist. Whereas the lower velocity clouds clearly 
manifested themselves in the HI emission measurements, the higher velocity
clouds were not detected. To illustrate this point, we show in Fig 1 the
optical absorption features and the HI emission profiles towards the
star HD 219188. It may be seen that there is no counterpart in HI
emission to the higher velocity optical absorption feature.
\begin{figure}
\begin{center}
\epsfig{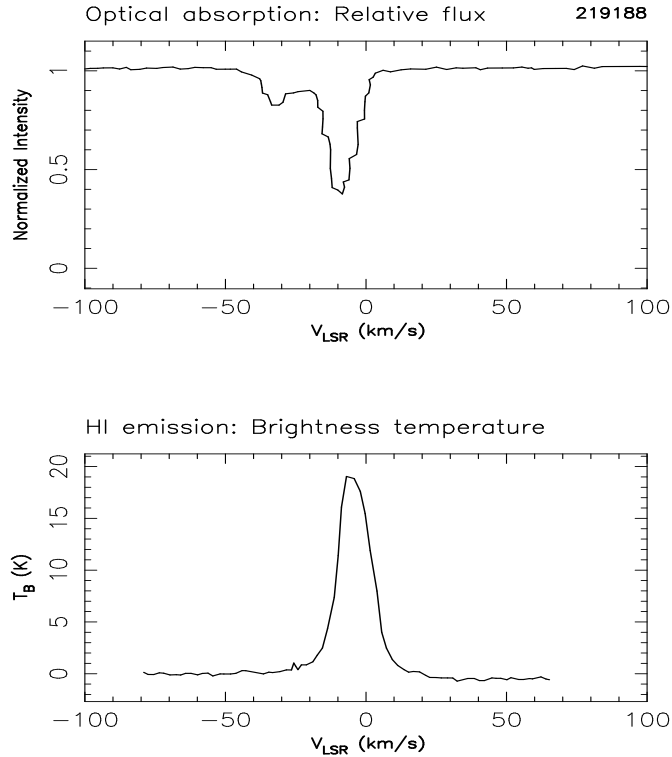}
\caption{Comparison of optical absorption (top) from CaII towards HD 219188
(Sembach, Danks and Savage, 1993) and HI
emission (Habing 1968) in the same direction. The absorption feature at higher
velocity is missing in the HI profile.}
\end{center}
\end{figure}

In the next two sections (sections 2 and 3) we will discuss the results
of our absorption survey presented in paper I. We shall classify the
HI absorption features into two broad classes, viz., ``low velocity''
and ``high velocity''. It is of course difficult to define a sharp
dividing line between ``low'' and ``high'' velocities. But based upon
earlier analyses of NaI to CaII ratios, as well as HI emission measurements,
we adopt a velocity of the order of 10 \kms as the dividing line. The
main conclusions from our study are summarized in section 4. Section 5
is devoted to a detailed discussion of the nature and origin of the high
velocity clouds.
\section{The low velocity clouds}
In this section we discuss the low velocity absorption features (i.e,
$ v \lsim 10$\kms). To recall from paper I, we detected HI absorption
in all but 4 of the 24 fields we looked at. We will discuss these four unusual
fields at the end of the next section.  

\subsection{Coincident absorption features}
All the lines of sight in our sample show optical absorption from
CaII at both low and high velocities. In most of the fields where
we have detected HI absorption, they occur at roughly the same
velocities as the optical absorption lines for $ v \lsim 10$\kms.
Not surprisingly, the HI absorption features have one to one correspondence
with the HI emission features in the fields for which earlier emission
measurements exist. In Table 1 we have listed all the fields with
``matching'' velocities in optical absorption and HI emission and
absorption.
\begin{table}
\begin{center}
\begin{tabular}{|l|c|c|l|c|l|}\hline
Field & V$_{lsr}$(opt) & V$_{lsr}$(HI)& $\tau$ & $\Delta V$ & T$_s$ \\ \hline\hline
{} & \kms  & \kms &  & \kms & K     \\ \hline
14143-&  -10.3& $-$11.2& 0.18& 11.8 & 283  \\
14134 & & & & & \\
14818&  -6.6&  -3.7 & 0.43 & 13.7 & 143  \\
21278&  -0.2&  2.8& 0.18 & 5.1 & 250  \\
21291&   -7.5& 7.0 & 0.60 & 6.2 & 180 \\
24912&  4.7&  4.3& 0.80 & 14.9 & 73  \\
25558&  10.1&  8.1& 1.13 & 5.6 & 73  \\ 
34816&  4.1&  6.0& 1.50 & 3.2 & 45   \\
41335&   0.2&  1.0& 0.22 & 7.2 & 342  \\
42087&   10.2&  12.4& 1.26 & 3.5 & 211   \\
141637&  0.0&  0.5 & 1.60 & 8.0 & NA  \\
148184&  2.2&  3.4& 4.90 & 3.4 & 50  \\
156110&  0.4&  2.2& 0.28 & 7.5 & 102  \\
159176&  -22.5&  $-$20.8&1.15 & 15.3 & NA   \\
166937&  5.9 &  5.4 & 1.70 & 5.7 & NA   \\
175754&  5.9&  6.8& 0.35 & 17.7 & 135  \\
199478&  -2.1, 8.7& 3.8$^+$ & 0.60 & 11.4 & 121  \\
212978&  0.6&  0.3&  0.20 & 9.0 & NA  \\
214680&  0.1&  1.4& * & 2.1 & NA  \\ \hline
\end{tabular}
\vspace{.5cm}
\caption[Summary of coincident low velocities.]{Summary of coincident low velocities: Column 1 lists the HD number for the field.
Columns 2 and 3 give the LSR velocities for the ``matching'' optical and HI absorption features
respectively. Column 4 and 5 lists the optical depth and width for the HI absorption derived from
the fitted gaussian. Column 6 lists the spin temperature T$_s$. These values have been derived by
using the optical depth from our observations along with the emission brightness temperatures
from Habing (1968). Where this was not available we used brightness temperatures from the
Leiden-Green Bank survey (Burton, 1985). Those fields where emission measurements were not
available are marked NA in column 6.\\+ The feature is a blend\\$*$ The feature is saturated}
\end{center}
\end{table}

For illustration we have shown in the upper panels of Fig. 2 the HI absorption
spectra (optical depth) in 3 fields; HI optical depth is plotted as a function
of V$_{\rm LSR}$. The arrows indicate the velocities of the optical absorption
lines. For comparison, we have shown in the lower panels the HI emission
in these 3 fields: the data obtained by Habing (1968) has been digitized
and re-plotted.  The first field contains the star HD 34816. The optical
spectrum obtained by Adams (1949) towards this star shows CaII absorption
at $-$14.0 and $+$ 4.1 \kms. The HI absorption profile shows a prominent
feature at 6 \kms (we have obtained spectra towards 2 radio sources in this
field but only one of them is shown here). As may be seen from the lower
panel, there is a corresponding emission feature at this velocity. As
mentioned in paper I, absorption features in the optical and HI spectra
may be taken to be at ``matching'' velocities provided they are within
$\sim$3 \kms of one another (this window is to account for blending effects
in the optical spectra and different corrections adopted for solar motion).
In view of this one may conclude that the optical absorption at $+$ 4.1 \kms
and HI emission and absorption at 6 \kms arise in the same cloud, even though the 
radio source is 20' away from the star.

The second panel pertains to the field containing the star HD 42087. 
In this field also we have two radio sources within the primary beam, and
the spectrum towards one of them is shown. The absorption features are
clearly seen at 4.4 \kms and 12.4 \kms, with the latter being much
stronger. The HI emission spectrum (shown in the panel below) shows a
broad peak centred at $\sim$ 15 \kms. The absorption feature at 12.4 \kms
may be taken to be the counterpart of the optical absorption
at 10.2 \kms. There is no HI absorption at negative velocities corresponding
to the other optical absorption lines indicated by the arrows. This may be
due to the fact that the 2 radio sources in the field are 32' and 42' respectively
from the star in question. Given that the star is at a distance of 1.3 kpc
(paper I, Table 1) it is conceivable that we are not sampling all the gas seen
in optical absorption. 

The spectra towards HD 148184 is shown in panel 3. Again there is good
agreement between the HI spectrum and the optical spectrum as far as the
lower velocity optical absorption is concerned. As in the previous two cases
there is no counterpart of the higher velocity optical absorption in the
HI spectrum. These two examples will suffice to illustrate the general
trend in Table 1 viz., there is reasonably good agreement at low velocities
($v \lsim 10$ \kms) between the optical absorption features and the HI
spectra.
\begin{figure}
\begin{center}
\vspace{6.0cm}
\mbox{
\epsfig{file=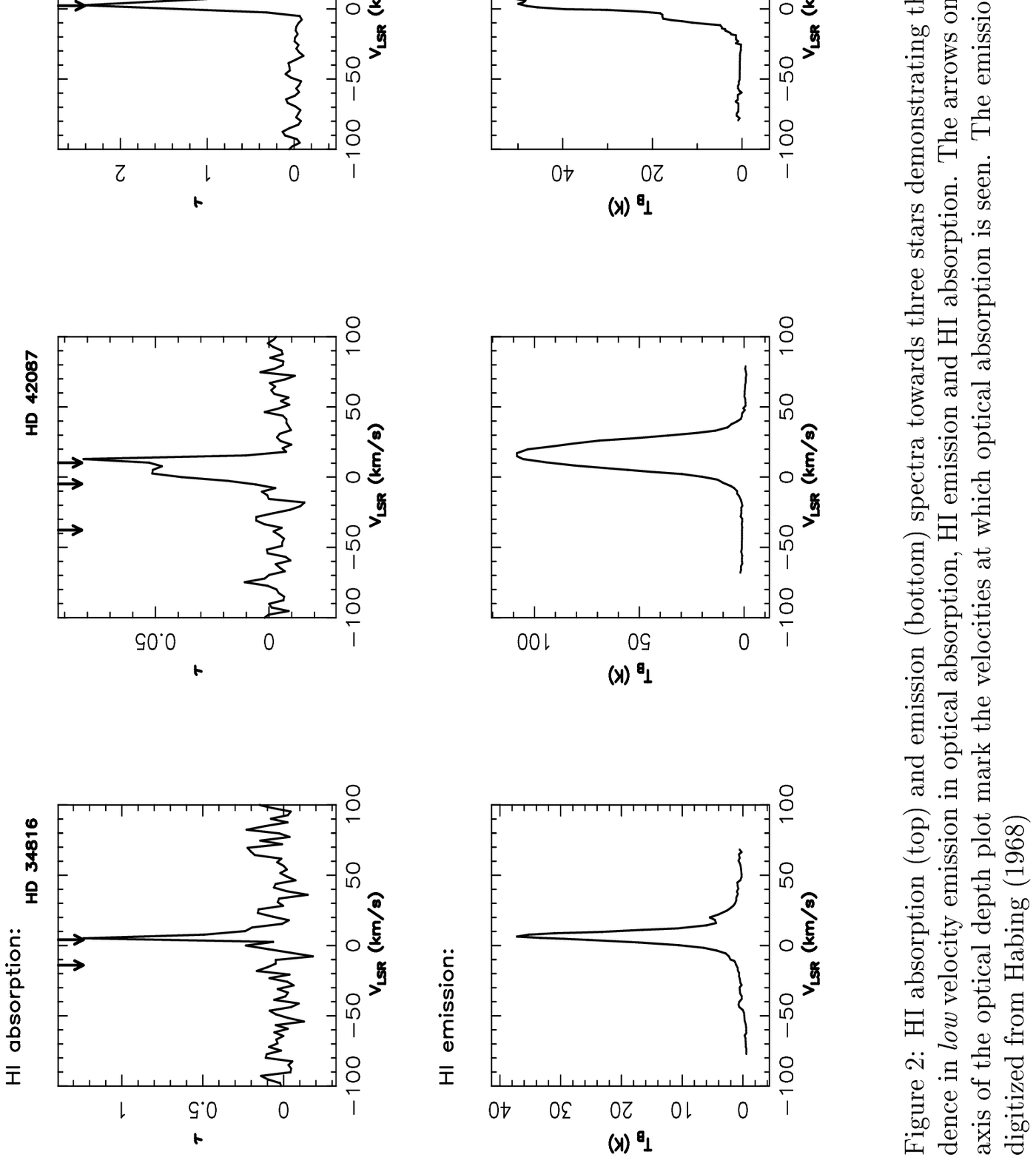,width=13cm,angle=90}}
\end{center}
\end{figure}

Returning to Table 1, we have listed the derived optical depths in column
4 and the velocity width of the HI absorption in column 5. In the last
column we have given the derived {\it spin temperatures}. These are
obtained by supplementing our HI absorption measurements with brightness
temperatures derived from emission measurements (mostly from Habing, 1968;
and in some cases from Burton, 1985). To the best of our knowledge, this
is the first direct determination of the temperatures of the interstellar
clouds seen in optical absorption. To be precise, the temperature derived
by us is the spin temperature which may be taken to be an approximate
measure of the kinetic temperature.

The correspondence between the optical absorption features and the
HI absorption suggest that one is sampling the same clouds in both
cases. The derived spin temperatures and velocity widths are consistent 
with these clouds belonging to the same population as the standard
cold diffuse clouds in the raisin-pudding model of the ISM.
While it is conceivable that at low velocities one may merely
be sampling the local gas (regardless of direction), statistical
tests carried out by Habing (1969) suggest that this is unlikely.
\subsection{Non-coincident absorption features}
As we have already encountered in the case of HD 42087 (see Fig. 2), sometimes
there is a mismatch between optical and radio spectrum even at low velocities.
We mention two specific cases here.
\subsubsection*{HD 37742}
The HI absorption spectrum obtained by us (Fig. 3) shows a deep absorption
feature at 9.5 \kms. There are no other absorption features down to an optical
depth limit of 0.03. The optical absorption features are at 3.6 and $-$21 \kms.
Since the radio source is only 12' away from the star, given the distance
estimate of 500 pc to the star the discrepancy between the optical spectrum
and the HI absorption spectrum is significant and intriguing. While the
gas seen strongly absorbing in HI could be located beyond the star, one
is left wondering as to why one does not see the low velocity cloud seen
in optical absorption.
\begin{figure}
\addtocounter{figure}{1}
\begin{center}
\mbox{
\epsfig{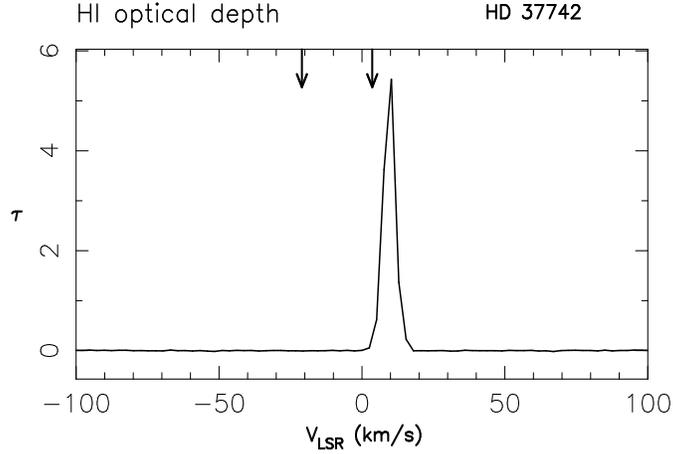}}
\end{center}
\caption{Spectrum towards
HD 37742. The arrows on the velocity axis
mark the velocities at which optical absorption is seen.}
\end{figure}
\subsubsection*{HD 119608}
The optical absorption spectrum towards this star obtained by
M\"{u}nch and Zirrin (1961) shows 2 minima at 1.3 and 22.4 \kms. We
have obtained HI absorption towards 2 strong radio sources in this
field both within 15' of the star. Both show a deep absorption feature
at $-$5.4 \kms (Fig. 4). This agrees with the HI emission feature
shown in the lower panel. However the HI emission spectrum also shows
a broad feature peaking at $\sim$ 20 \kms. This is also seen
in the more recent measurement of Danly etal (1992). If this feature
is indeed to be identified with the optical absorption at 22.4 \kms, 
then this represents an interesting case where the gas in the line of
sight to the star causing optical absorption manifests itself in
HI emission but not absorption. This could happen for example if the
spin temperature of this gas is sufficiently high as to make the HI
optical depth below our detection limit. HD 119608 is a high latitude
star and one is presumably sampling the halo gas, and warrants a deeper
absorption study.
\begin{figure}
\begin{center}
\epsfig{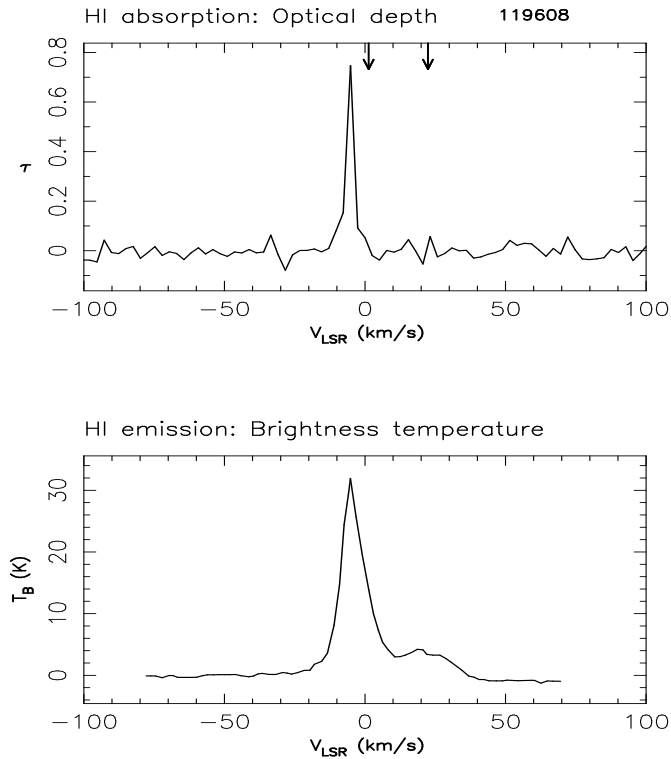}
\end{center}
\caption{HI absorption (top) and emission (bottom) spectra towards
HD 119608. The arrows 
mark the velocities at which optical absorption is seen. The emission spectrum
is from Habing (1968).}
\end{figure}
\section{The high velocity clouds}
Although in the previous section, we were primarily concerned with
establishing the correspondence between the optical absorption features 
and the HI absorption spectra at low velocities, we did have occasion to
comment on the {\it absence of HI absorption from the high velocity
clouds} [The high velocity clouds we are discussing are those that populate
the tail of the velocity distribution obtained by Blaauw (1952) and not
those that are commonly referred to as HVCs in the literature]. 
It turns out that in all but 4 cases, we fail to detect HI absorption
at velocities corresponding to the high velocity \mbox{($\gsim$ 10 \kms)}
optical absorption lines. To illustrate this generic trend, we have
shown some additional examples in Fig. 5. {\it It may be seen in the
figure that the high velocity optical absorption features (indicated
by arrows) are not seen in the HI emission spectra either}. A discussion
of this will form the major part of section 5.

\begin{figure}
\begin{center}
\vspace{6.0cm}
\epsfig{file=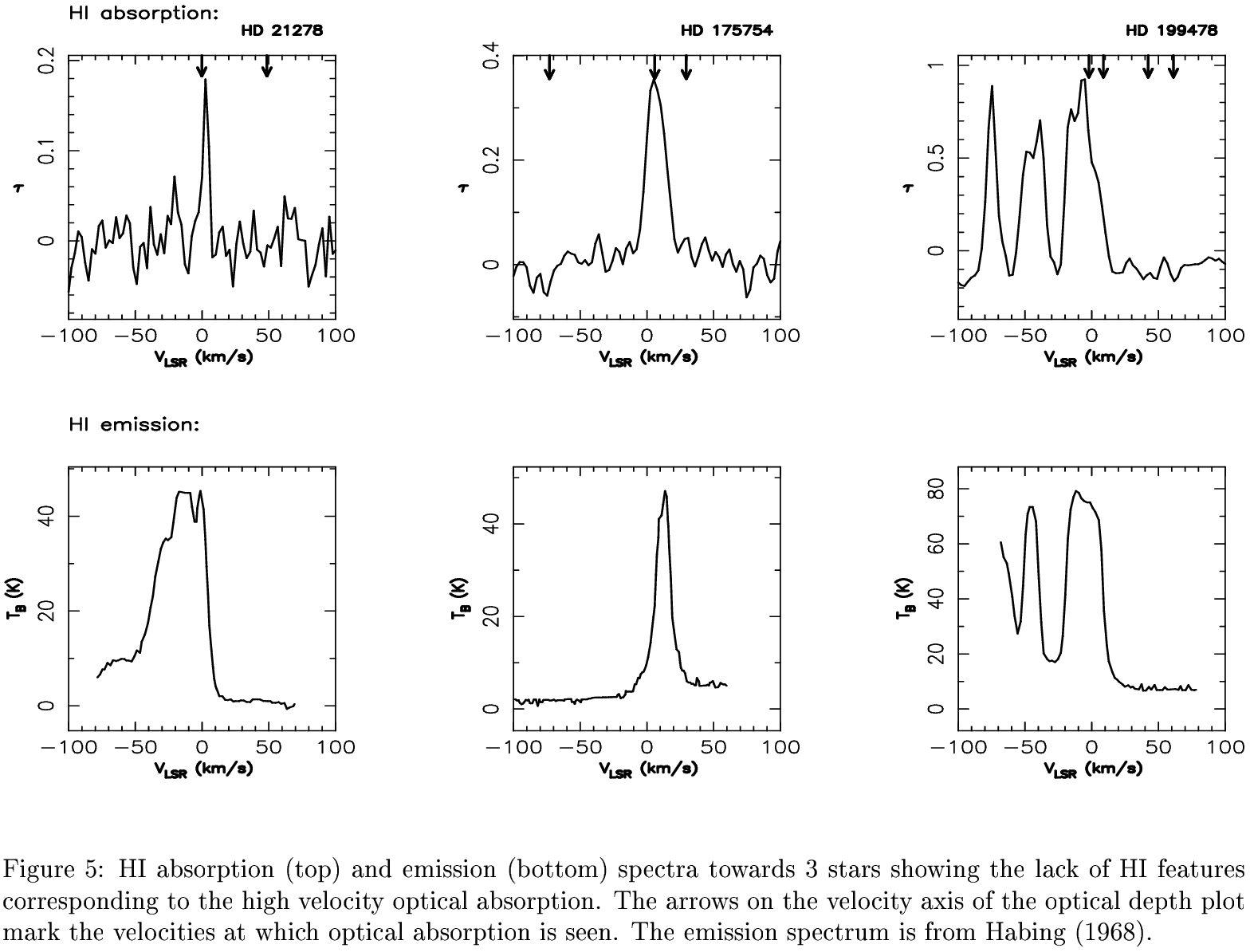,width=13cm,angle=90}
\end{center}
\end{figure}
In only 4 out of the 24 lines of sight have we detected HI absorption at
velocities $\gsim$ 10 \kms and which clearly correspond to the optical absorption
lines. We discuss these below.
\subsection{Coincident absorption features at high velocities}
\subsubsection*{HD 14134, HD 14143}
These two stars (in the same field)
are members of the h and $chi$ Persei clusters (M\"{u}nch 1957). 
There were 3 radio sources within our primary beam (all within 10' of
the star). The HI absorption features towards one of them is shown in
Fig. 6. The prominent high velocity absorption features towards the
3 sources are at $-$52.8, $-$50.3 and $-$46.1 \kms, respectively (Table 1
of paper I). These should be compared with the optical absorption features
towards the 2 stars in question which are at $-$46.8 and $-$50.8 \kms.
Thus there is reasonable coincidence between the optical and HI data.
Nevertheless we wish to now point out that the high velocity may not
represent random motion but rather systematic motion. 
\begin{figure*}
\addtocounter{figure}{1}
\begin{center}
\epsfig{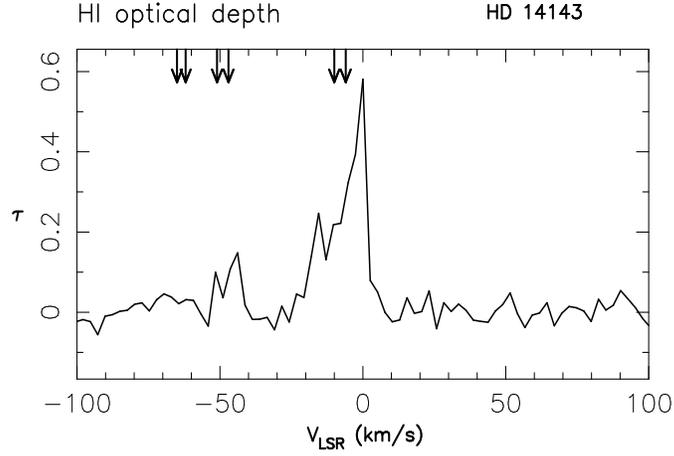}
\end{center}
\caption{HI absorption spectra towards
HD 14143.
The arrows mark the velocities at which optical absorption is seen.}
\end{figure*}
While interpreting his pioneering observations M\"{u}nch (1957)
attributed the high velocity features in the optical spectra to
anomalous motions in the Perseus arm. Since then it has generally
been accepted that there are streaming motions in the Perseus arm
with velocities ranging from $-$10 to $-$30 \kms (Blaauw and Tolbert
1966; Brand and Blitz 1993). For the sake of completeness we have listed
in Table 2 the spin temperature of the gas derived by us by combining our
measurements with existing HI emission measurements.
\begin{table}[t]
\begin{center}
\begin{tabular}{|l|l|l|l|l|l|}\hline
Field & V$_{lsr}$(optical) & V$_{lsr}$(HI)& $\tau$ & $\Delta V$ & T$_s$ \\ \hline\hline
{} & \kms  & \kms &  & \kms & K     \\ \hline
14143-& $-$50.8& $-$50.3 & 0.30 & 10.0 & 96  \\
14134 & & & & &  \\
21291&  $-$34.0&  $-$31.2&  0.40 & 3.3 & 181  \\
159176&  $-$22.5&  $-$20.8& 1.15 & 15.2 & NA  \\ \hline
\end{tabular}
\vspace{.5cm}
\caption[Summary of coincident high velocities.]{Summary of coincident high velocities: Column 1 gives the HD number of the field.
Column 2 lists the {\em high} velocity optical absorption seen towards the star. Column 3 is the ``matching'' HI absorption.
Columns 4 and 5 give the fitted optical depth and width of the HI absorption
features. The spin temperature is listed in column 6. As in the  for low velocity features, we
have used Habing(1968) and the Leiden-Green Bank survey for the emission temperatures needed to
compute the spin temperature from the optical depth.}

\end{center}
\end{table}

\subsubsection*{HD 21291}
The spectrum towards this star near the Perseus arm has a prominent Na D line 
at a velocity of $-$34 \kms (M\"{u}nch, 1957).
The HI absorption spectrum shows a feature at $-$31.2 \kms.
The contribution to radial velocity from Galactic rotation 
can only be $\sim$ 10 \kms, thus indicating 
significant peculiar motion of the gas. 
In our opinion one must attribute this to streaming motion of the gas
as in the case discussed above. HI emission clearly shows spatially extended
gas covering the longitude range from $\sim$136$^{\circ}$ to 141$^{\circ}$ at
the velocity of interest.
This strengthens the conclusion that one must not attribute the observed
velocity to random motions.

\subsubsection*{HD 159176}
There is pronounced optical absorption at $-$ 22.5 \kms which
might be identified with the HI absorption seen by us at $-$20.8
\kms. Given the longitude of 356$^{\circ}$, it is difficult to attribute
this to Galactic rotation. The measured velocity must correspond 
either to random velocity or systematic motion. 
\subsubsection*{HD 166937}
In the case of this star, there is no strict coincidence (within
3 \kms) between the HI and optical absorption at high velocities.
However we see two HI absorption features at velocities close to
and straddling the optical feature at 41.1 \kms, so
we include this field in our list of high velocity coincidences.
It may be noted that this star is also close to the Galactic center
direction.

\begin{figure}
\begin{center}
\mbox{
\epsfig{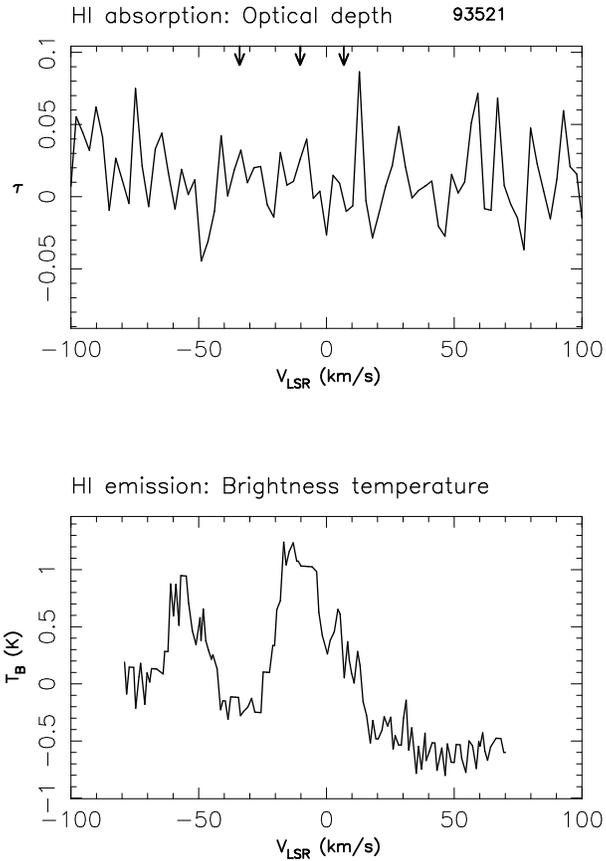}}
\end{center}
\caption{HI absorption (top) and emission (bottom) spectra towards
HD 93521. The arrows plot mark the velocities at which optical absorption
 is seen. The emission spectrum
is from Habing (1968).}
\end{figure}
\section{Summary of results}
In the preceding section we described
 the first attempt to directly compare
HI absorption with optical absorption features arising in the ISM. We summarize
below the main results:

\begin{itemize}
\item[1.] HI absorption measurements were carried out towards 24 fields. Each field
has existing optical absorption spectra towards a bright star.
In 20 of these fields we detected HI absorption features.
\item[2.] In all but 4 of these 20 fields, the HI
absorption features at
low velocities ($<$ 10 km s$^{-1}$) correspond to the optical absorption lines.
\item[3.] In most cases there is also corresponding HI emission.
\item[4.] The
spin temperatures derived by us for the low velocity gas is consistent with 
the standard values for the 
cold diffuse HI clouds.
\item[5.] This lends strong support to the hypothesis that (at least)
the low velocity clouds seen in optical absorption belong to the same
population as those sampled in extensive HI studies.
\item[6.] In 20 out of 24 fields surveyed, we did not detect any HI
absorption corresponding to optical absorption at high velocities
($v > 10$\kms). It is unlikely that in all cases this is due to our
line of sight not sampling the gas seen in optical absorption; in several
cases the line of sight to the radio sources would have sampled this gas
even if its linear size was of the order of 1 pc.
\item[7.] Curiously, the early emission measurements also failed to
detect HI gas at high velocities (in these same lines of sight). Given the
size of the telescopes used, beam dilution could have accounted for the
non-detection if the gas was ``clumpy''. Our absorption measurements rule
this out as a generic explanation. More recent and sensitive measurements
indicate that the high velocity gas has much smaller column density than
the low velocity gas (N(HI) $<$10$^{\rm 18}$ cm$^{-2}$). If this is the
case, then it is not difficult to reconcile why we do not see it in absorption
since our sensitivity limit was $\tau \gsim$0.1. But then the correlation
between high velocity and low column density would have to be explained.
We venture to offer some suggestions in the next section.

\item[8.] {\it Fields with no HI absorption}: We wish to record that in the fields
containing the stars HD 38666, HD 93521, HD 205637 and HD 220172 {\it we
did not detect any HI absorption - even at low velocities.} These are
all high latitude stars. HI emission spectra also show only weak 
features. For illustration we show in Fig. 7 the HI absorption and
emission spectrum towards HD 93521. From very detailed investigations -
the case of HD 93521 is a good example - it has been concluded that most of the
optical absorption arises from warm gas in the halo (detailed references
may be found in Spitzer and Fitzpatrick 1993 and Welty, Morton and Hobbs
1996). The fact that we do not see HI absorption is consistent with the
interpretation of this gas being warm. Weak HI emission indicates low 
column density also.
\end{itemize}

\section{Discussion}
As we have already argued, our absorption measurements, taken
together with earlier emission measurements, establishes that the
low velocity clouds seen in optical absorption are to be
identified with the standard HI clouds - their column densities
and spin temperatures match.

But the true nature of the high velocity clouds seen in
optical absorption is still unclear. There are two questions
to be addressed: (1) Do the high velocity clouds belong
to a different population, and (2) is there a causal connection
between their higher velocities and lower column densities?
We wish to address these two questions below.

An unambiguos indication that the high velocity clouds may
have very different properties compared to their low velocity
counterparts comes from an HI absorption study towards the
Galactic center (Radhakrishnan and Sarma 1980). Given the 
statistics of clouds derived from optical studies (8 to
12 per kpc), if the high velocity clouds had optical depths
comparable to the low velocity clouds, then an absorption
experimant towards the Galactic center should straightaway
reveal a velocity distribution similar to the one derived by
Blaauw (1952) from Adams' data. The velocity distribution 
derived by Radhakrishnan and Sarma from precisely such a study
did not reveal a pronounced high velocity tail. The velocity
dispersion of 5 \kms derived by them was in good agreement with
the low velocity component of Blaauw's distribution. Instead
of a pronounced high velocity tail seen in optical and UV studies, 
there was at best a hint of a high velocity population of very
weakly absorbing clouds. Even this conclusion has remained
controversial (Schwarz, Ekers and Goss 1982).

As for the possible correlation between higher velocities of
clouds and lower column densities, fairly conclusive
evidence comes from UV absorption studies. Since
the UV absorption lines have larger oscillator strengths,
they can be used to probe smaller column densities than is
possible with optical absorption lines.
The analysis of Hobbs (1984) seems to confirm this
expectation - in several lines of sight there is more
high velocity UV absorption features than in the optical.
A more direct inference can be drawn from the work of
Martin and York (1982). For the two lines of sight they
studied, there is a clear indication of lower column
density (N(HI)) at higher velocities.

Over the years, three broad suggestions have been put
forward in an attempt to elucidate the nature of the high 
velocity clouds.

\subsubsection*{Circumstellar clouds}
According to an early suggestion due to Schl\"{u}ter, 
Schmidt and Stumpf (1953), the high velocity clouds
seen in optical absorption are to be identified with
circumstellar clouds. This was an attempt to explain the
predominance of {\it negative} velocities in the high velocity
absorption features. If the clouds in the vicinity of
massive stars are accelerated by the combined effect of
stellar winds and radiation from the stars, then in  an
absorption study against the stars one would detect only 
those clouds accelerated towards us. A few years later, Oort
and Spitzer (1955) developed the well known ``rocket mechanism``
in which the UV radiation from the star ionizes the near side
of the cloud resulting in ablation and consequent acceleration
of the cloud. This mechanism will naturally result in the
higher velocity clouds having smaller mass and therefore smaller
column density. The difficulty with this mechanism, however, is 
that one will have to invoke another mechanism to explain the
large {\it positive} velocities which are also seen in absorption
studies. In view of this we will not dwell any further on 
this scenario.

\subsubsection*{Relic SNRs}
An alternative scenario was advanced by Siluk and Silk (1974).
Their suggestion was that the high velocity optical absorption
features arise in very old supernova remnants (SNRs) which have
lost their identity in the ISM. Their primary objective in
advancing this scenario was to explain the high velocity tail
of the velocity distribution of optical absorption features.
The point was that if the absorption features arise not in
interstellar clouds but in SNRs in their very late stages of
evolution, then it would result in a power law distribution
of velocities; such a distribution according to them provided
a good fit to the observations.

While this suggestion is quite attractive, it suffers from two
drawbacks: (1) The early studies on the evolution of
SNRs predicted the formation of very dense shells beyond the radiative
phase. Such compressed shells were essential to  
explain the observed absorption features and the derived
column densities. However, more recent studies which take
into account the effects of the compressed magnetic field and
cosmic ray pressure in the shells suggest that either dense
shells do not form or if they do, do not last long enough
(Spitzer 1990; Slavin and Cox 1993).
(2) Given a supernova rate of one per $\sim$ 50 years
in the Galaxy, the statistics of absorption features requires 
that the SNRs enter the radiative phase (and as a consequence
develop dense shells) when they are still sufficiently small
so as not to overlap with one another. This would indeed be the
case if the intercloud medium into which the SNRs expand is
dense enough (n $\sim$ 0.1 cm$^{-3}$). But if a substantial
fraction of the ISM is occupied by low density hot gas
(n$\sim$0.003 cm$^{-3}$; T $\sim$ 5$\times$10$^5$K) such
as indicated by UV and soft X-ray observations then the supernova
bubbles are likely to intersect with one another and perhaps even
burst out of the disk of the Galaxy before developing dense
shells (Cowie and York 1978).
In view of these two drawbacks, we do not favour this suggestion.
\subsubsection*{Shocked clouds}
The third possibility is that the high velocity absorption features
do arise in interstellar clouds but which have been engulfed and
shocked by supernova blast waves. Indeed we feel that this is the
most plausible explanation for it has support from several quarters.
The earliest evidence that the high velocity gas may be ``shocked''
came from the Routly-Spitzer effect. The NaI/CaII ratio in the
fast clouds was lower (sometimes by several orders of magnitude)
than in the slow clouds. The variation in NaI/CaII ratio was
primarily attributed to the variable {\it gas phase abundance} of
calcium in these clouds. Due to its relatively high condensation
temperature calcium is likely to be trapped in grains. 
Spitzer has argued that the observed trend in NaI/CaII
ratio could be understood if the calcium is released back
into the gas phase in the high velocity clouds due to
sputtering. This is indeed what one would expect if the interstellar
cloud is hit by an external shock, which in turn drives a
shock into the clouds (Spitzer 1978). Supernova blast waves are
the most likely candidates.

Earlier in this section we referred to an HI absorption study by 
Radhakrishnan and Sarma towards the Galactic center. While they
did not find strong absorption at high velocities they did conclude
that there must be a population of weakly absorbing high velocity
clouds. Radhakrishnan and Srinivasan (1980)
examined this more closely and advanced the view that in order to
explain the optical depth profile centered at zero velocity one
had to invoke {\it two distinct velocity distributions}: a standard
narrow distribution with a velocity dispersion of $\sim$ 5 \kms,
and a second one with a much higher velocity dispersion of
$\sim$35 \kms. While arguing strongly for a high velocity tail, they
stressed that the latter distribution must consist of a population
of very weakly absorbing clouds. They went on to suggest that
this population of weakly absorbing clouds might be those that
have been shocked by expanding SNRs; the very process of acceleration
by SNRs might have resulted in significant loss of material and
heating of the clouds, leading to low HI optical depths.

To conclude this discussion we wish to briefly summarize the
expected life history of a cloud hit by a supernova blast
wave. The first consequence of a cloud being engulfed by
an expanding SNR is that a shock will be driven into the
cloud itself resulting in an eventual acceleration of the
cloud. The effect of this shock and a secondary shock propogating
in the reverse direction after the cloud has been overtaken
by the blast wave, is to compress and flatten the cloud. Eventually
various instabilities are likely to set in which
will fragment the cloud.

The detailed history of the cloud depends upon two important
timescales: the time taken for the cloud shock to cross the
cloud and the evolutionary timescale of the SNR. If the
former is much smaller than the latter, the cloud is likely
to be destroyed. However if the reverse is true, then the shocked cloud
will survive and be further accelerated as a consequence of the
viscous drag of the expanding hot interior. {\it Clouds accelerated in such 
a manner will however suffer substantial evaporation due to heat
conduction from the hot gas inside the SNR.} Partial fragmentation
could further reduce the size of the cloud. For detailed
calculation and discussion we refer to McKee and Cowie (1975),
Woodward (1976), McKee, Cowie and Ostriker (1978), Cowie, 
McKee and Ostriker (1981) and a more recent paper by Klein,
McKee and Woods (1995).

To summarize the above discussion, in our opinion the shocked cloud 
scenario has all the ingredients needed to explain the observational
trends. In particular it would explain why the high velocity
clouds seen so clearly in optical and UV absorption lines
do not manifest themselves in HI observations. But this observation
is predicated on the conjecture that the higher velocity clouds
are not only warmer, but have smaller column densities. There
is certainly an indication of this from optical and UV absorption
studies. To recall, in UV observations which are sensitive to
much smaller column densities than optical studies, the higher
velocity absorption features are more pronounced. But it would
be desirable to quantify the correlation between velocity and
column densities. Reliable column densities are difficult to 
obtain from optical observations because of blending of lines
and also depletion onto grains. The column densities derived from
UV observations are also uncertain because of the effects of
saturation of the lines. In view of these difficulties it would
be rewarding to do a more systematic and much more sensitive
HI absorption study, supplemented by emission studies.

\label{lastpage}
\end{document}